# Does Very Short Gamma Ray Bursts originate from Primordial Black Holes? Review

D.B. Cline[a] and S. Otwinowski[b]

[a]*UCLA, CA 90024, USA, Department of Physics and Astronomy, Box 951457*
*e-mail: dcline@physics.ucla.edu*
[b]*UCLA, CA 90024, USA, Department of Physics and Astronomy, Box 951457*
*CERN, CH-1211, Geneve 23, Switzertand, e-mail: stanislaw.otwinowski@cern.ch*

B. Czerny[c] and A. Janiuk[d]

[cd]*Nicolaus Copernicus Astronomical Center, Bartycka 18, 00-716 Warsaw, Poland,*
*e-mail: [c]bcz@camk.edu.pl, [d]agnes@camk.edu.pl*

DRAFT

**Abstract**

We present the state of current research of Very Short Gamma Ray Bursts (VSGRBs) from seven GRB detectors. We found that VSGRBs form distinct class of GRBs, which in our opinion, in most cases can originate from the evaporating Primordial Black Holes (PBHs).

Arguments supporting our opinion:

1. GRBs with time duration ($T_{90}$) < 100 ms form distinct class: VSGRBs.

2. We observe significant anisotropy in the galactic angular distribution of BATSE VSGRB events.

3. V/Vmax distribution for BATSE VSGRB events indicates the local distance production.

4. VSGBBs have more energetic γ-ray burst than other GRBs with longer duration (KONUS).

5. We observe small number of afterglows in SWIFT VSGRB sample (25%), in contrast with the noticeable afterglow frequency in SGRB sample (78%).

6. Time profile of rising part BATSE VSGRBs is in agreement with the evaporation PBH model.

## 1. Introduction

Very shortly after the Big Bang, pressure and temperature were extremely great. Under these conditions, simple fluctuation in the density of the matter could have resulted in creation of black holes. Primordial black holes (PBHs) would persist to the present [1], [11]. Hawking [2] showed that PBHs of mass less than ~ $5 \cdot 10^{14}$ g must have been evaporated by now.

One way to detect PBHs is by their Hawking radiation, when they evaporate ("explode") at present. The properties of PBH burst emission are model dependent and were estimated [3] for γ-ray burst in the range of tens of ms with luminosity ~ $10^{33}$ erg. If we supposed the phase transition in final state of evaporation – what means "explosion" with upper limit of particles` and γ` energy, then we should expect modification of the final state of Hawking evaporation process.

## 2. Time duration

BATSE

GRB from detector BATSE we divide into three classes according to their time duration ($T_{90}$): Long Gamma Ray Bursts (LGRBs) ($T_{90} > 1$ s); Short Gamma Ray Bursts (SGRBs) (1 s > $T_{90}$ > 0.1 s); and VSGRBs ($T_{90} \leq 0.1$ s). See Fig.1. We assume that the VSGRBs constitute a separate class of GRBs with log normal duration distribution and we fit the time distribution in Fig.1 with a three-population model. The fit is excellent but does not give significant evidence for a three-population model. The arrow [↓] shows position of VSGRBs.

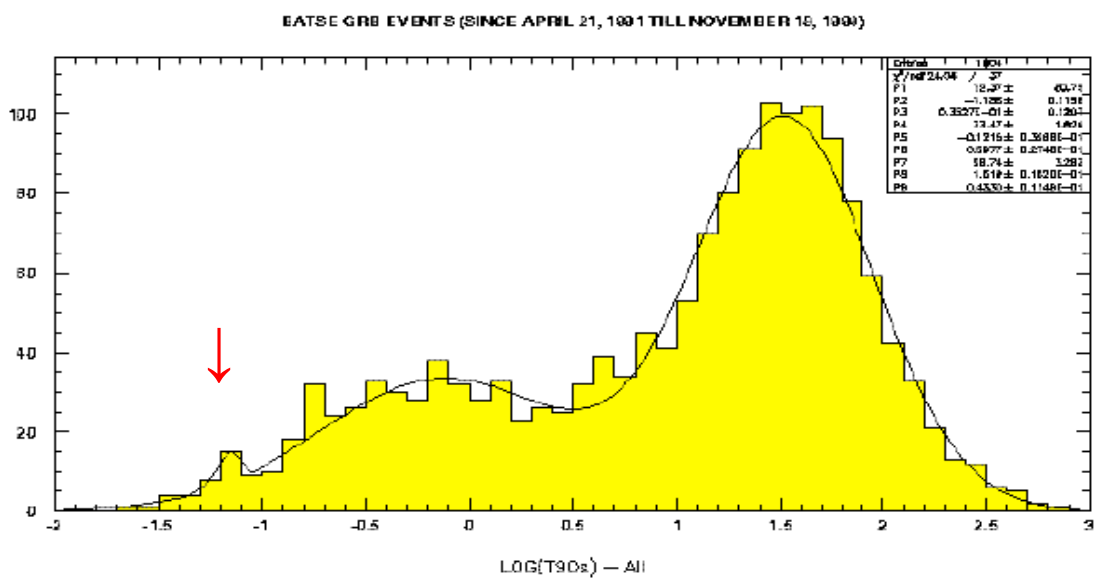

Fig.1. The time distribution $T_{90}$ for all GRB from BATSE detector [1].

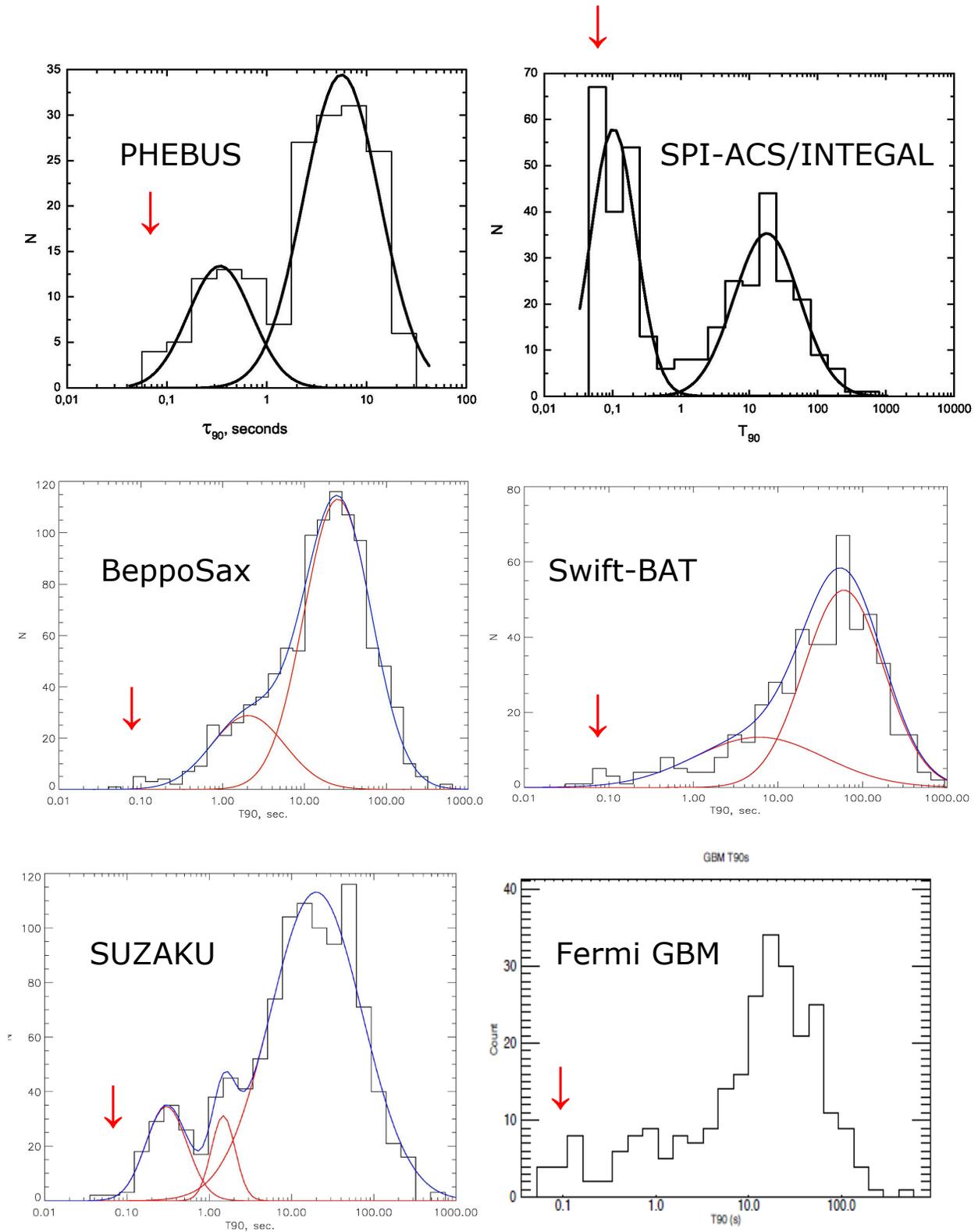

Fig.2. Duration distribution, $T_{90}$, for GRBs from PHEBUS, SPI-ACS/INTEGRAL, BeppoSax, Swift and SUZAKU detectors according to [4] and Fermi GBM GRB catalog. The arrow [↓] shows position of VSGRBs.

In Fig.1 and Fig.2 we present GRB duration distribution for 7 detectors. The distributions are different and they depend on different sensitivity (detector dimension), detected spectrum energy, the background dependence, trigger time length and other detector characteristics. There are well visible noticeable two or three groups: long and short time bursts, there is also possible to distinguish middle time bursts (energy extended). Looking at the shortest burst we can see some tail for each detector, which sticking out of the log normal fitted curves. That is not statistically significant but it is observed for **every** detector even if trigger minimum length reduces probability detection of shortest burst! In total it seems to be statistically significant. Additionally the energetic analysis of KONUS short and very short bursts mark out VSGRB group – see Fig.8. Also the non isotropic events show excess for GRBs with duration shorter than 0.1 s, look at Fig.4. All this suggests that we really observe distinct group of VSGRBs with $T_{90} \leq 0.1$ s.

### 3. Anisotropy

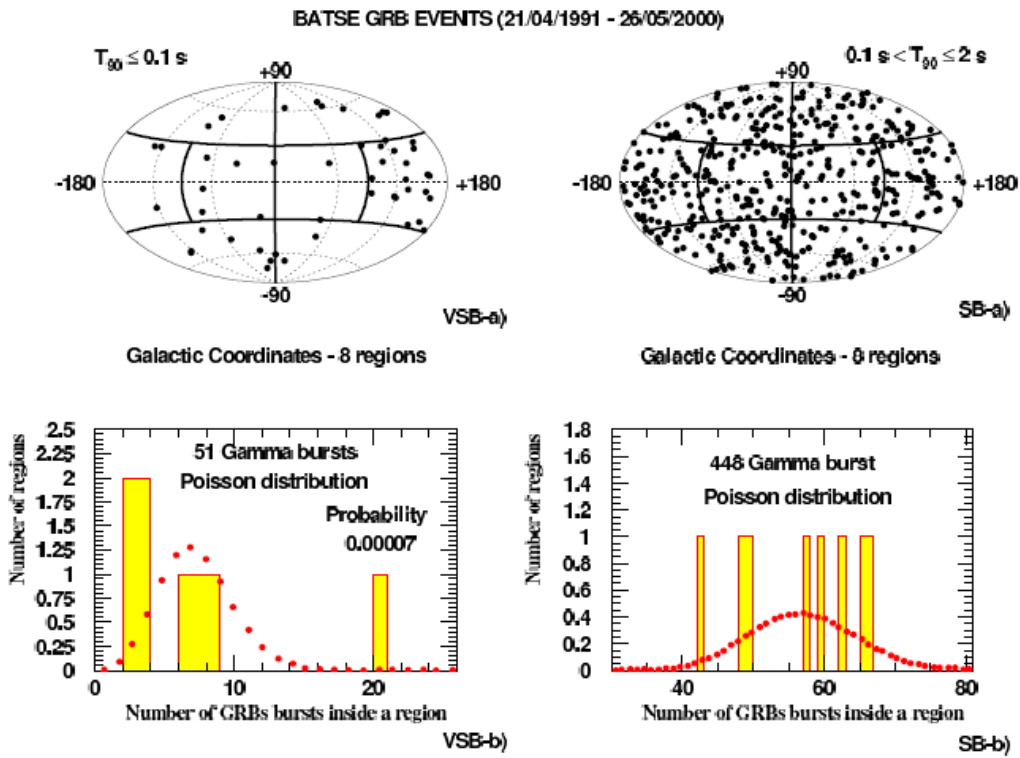

Fig.3. Angular distribution of the GRB in Galactic Coordinates and the corresponding histograms, in comparison with Poisson distribution predictions for Very Short Burst (VSB) and Short Burst (SB) $T_{90}$ ranges (full circles) [5].

The angular distribution of VSGRB and SGRB in Galactic Coordinates is shown in Fig. 3. The sky was divided into 8 equal regions. In the case of isotropic distribution the number of bursts in each region should be described by the Poisson distribution. The histograms of such values are shown in Fig. 3. The probability of any number of bursts in a single zone multiplied by the number of zones (8) is shown with full circles. For 448 SB (0.1 s < $T_{90}$ < 2 s) we see conformity with isotropic distribution, but for VSB ($T_{90} \leq 0.1$ s) the observed distribution is strongly improbable to be isotropic. The number of bursts in one of the regions (roughly in the direction of the Galactic Anticenter) is 20, which is much higher than the expected average of 51/8. The probability to find twenty or more events (from the total number of fifty one) in the region of 1/8 area is 0.00007. This result argues for other explanation than the statistical fluctuation. Background in the direction of the Galactic Center is $12500 \pm 1000$ counts·s$^{-1}$, while the mean level of the background outside this region is $13800 \pm 1300$ counts·s$^{-1}$, and the total number of SB in these regions is slightly lower than the expected average (but within the expected error). Therefore, background anisotropy cannot be responsible for the observed distribution of the VSB across the sky [5].

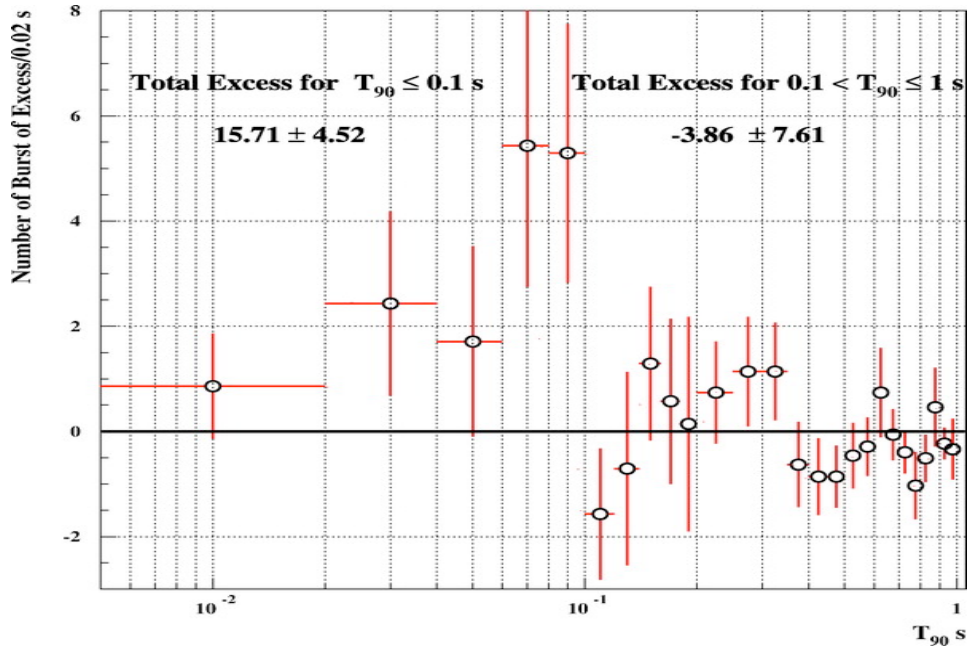

Fig.4. BATSE GRB events (1991 Apr 21 – 2000 May 26). Excess in GRBs inside the chosen region (see Fig. 3) (30º < b < -30º; 90º < l < 180º) as a function of $T_{90}$ [5].

In Fig.4 we show the distribution of the number of the GRBs excess as a function of $T_{90}$. The excess from the isotropic distribution is calculated as a difference between the GRBs number in the chosen region and the sum of all GRBs number in other regions divided by seven. This study shows that the incompatibility with isotropic distribution is seen only for GRBs with $T_{90} \leq 0.1$ s. We observe the total excess of $15.71 \pm 4.52$ bursts in this region [5].

The BATSE VSGRB angular Galactic coordinate distribution was reanalyzed using factorial moments and cumulants analysis [6]. Authors present VSGRB angular distribution with very suggestive method, which consists in presenting GRB coordinate point to 10º, 25º and 40º radius cone, see Fig.5. The detailed first four factorial moments analysis gives as a result the probability $< 3·10^{-5}$ for the chance of such fluctuation from uniform distribution. This is in agreement with our earlier, simple estimation: $7·10^{-5}$, see Fig.3, VSB-b. It means the effect itself is on about $4\sigma$ level.

The authors analyze, with cumulants method, up to 5<sup>th</sup> order, if there is any structure within this cluster. The result is consistent with the lack of any genuine correlation of 4<sup>th</sup> and 5<sup>th</sup> order and suggests that the there are few smaller groups of multiplicity about of 2-3. This result is "scale-free".

The anisotropy of VSGRBs suggests the question if we observe any correlation with other phenomenon.

In [7] the possible correlations of VSGRBs with CMB, cosmic rays and particular kind of astronomical objects (Polars) is shown. These correlations show particularity that in Quadrant 2 some astronomical indicators are stronger [7],[8]. It is in agreement with our observation of the VSGRB concentration in this region. Is it by chance? We cannot exclude that such concentration of matter was relic from the time of the Big Bang.

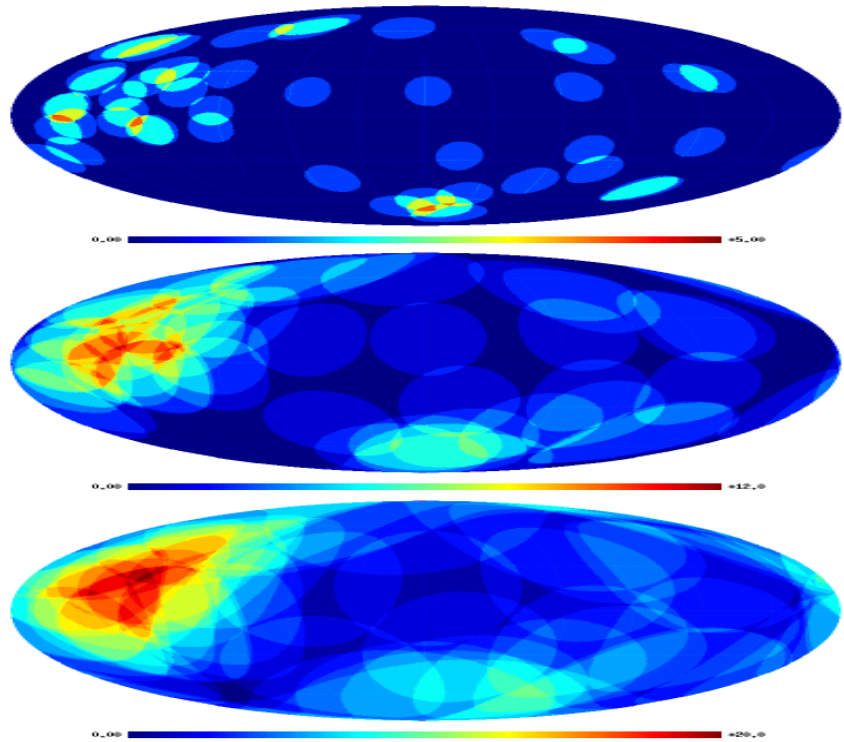

Fig.5. Angular distribution of the BATSE VSGRB events in Galactic Coordinates within 10º (top), 25º (middle) and 40º (bottom) radius cone around each event [6].

## 4. V/Vmax distribution

We have also reanalyzed the overall radial distribution of the VSGRBs and the SGRB using the standard ‹V/V$_{max}$› test [3]. We used the C$_{max}$/C$_{min}$ table from BATSE catalog as an input to V/V$_{max}$ calculations: V/V$_{max}$ = (C$_{max}$/C$_{min}$)$^{-3/2}$. The results are ‹ V/V$_{max}$ › = 0.52±0.05 and 0.36±0.02 for VSGRBs and SGRBs respectively [5], see Fig.6. It means that SGRBs appear to come from cosmological distances, but no cosmological effects are seen in the distribution of VSGRBs. We have one more confirmation of the difference between both groups of events.

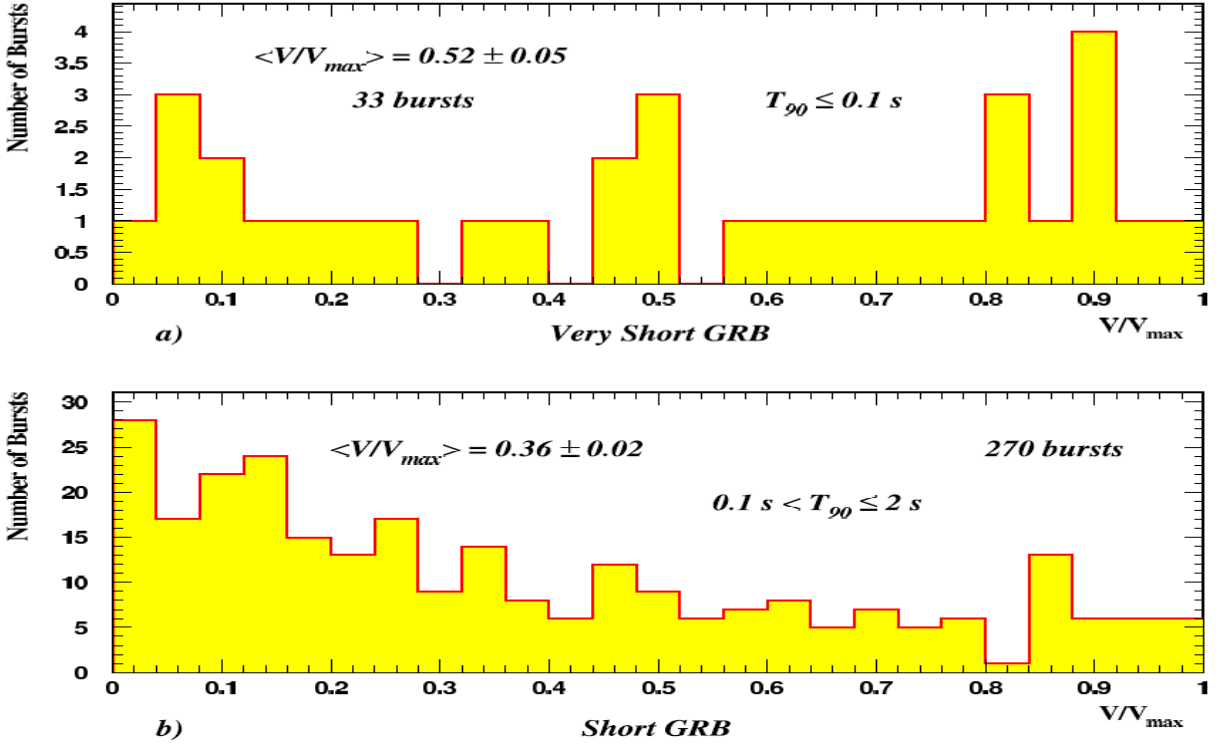

Fig.6. Distribution of the V/$V_{max}$ for BATSE events (1991 April 21 – 2000 May 26) [5].

**5. VSGBBs have more energetic γ-ray bursts than other GRBs with longer duration**

In Fig.7 we compare the mean energy of SGRB and VSGRB for KONUS events. Nominal energy range of γ-ray measurements in KONUS covers the interval from 12 keV up to 10 MeV. We observe that in the MeV region the spectra of VSGRB are significantly harder than the spectra of SGRB. The spectrum starts to be flatter above 3 MeV and the effect in the case of VSGRB is stronger [5].

To follow up in Fig. 8 we study all SGRB events in KONUS data and select burst with the mean energy ‹Eγ› > 90 keV. We construct a histogram of burst numbers as a function of their duration. For comparison, we show similar histogram made for SGRs with mean energies ‹Eγ› < 90 keV, and scale it down to the distribution of harder bursts. Comparing these two distributions we see very strong clumping of hard bursts at very short durations. In this histogram in time interval $T_{90} < 0.1$ s we expect one event from softer bursts distribution and found ten hard events, which is extremely unlikely (probability of such fluctuation is < $10^{-10}$). That indicates again some new physics origin of the bulk of the VSGRB data [9].

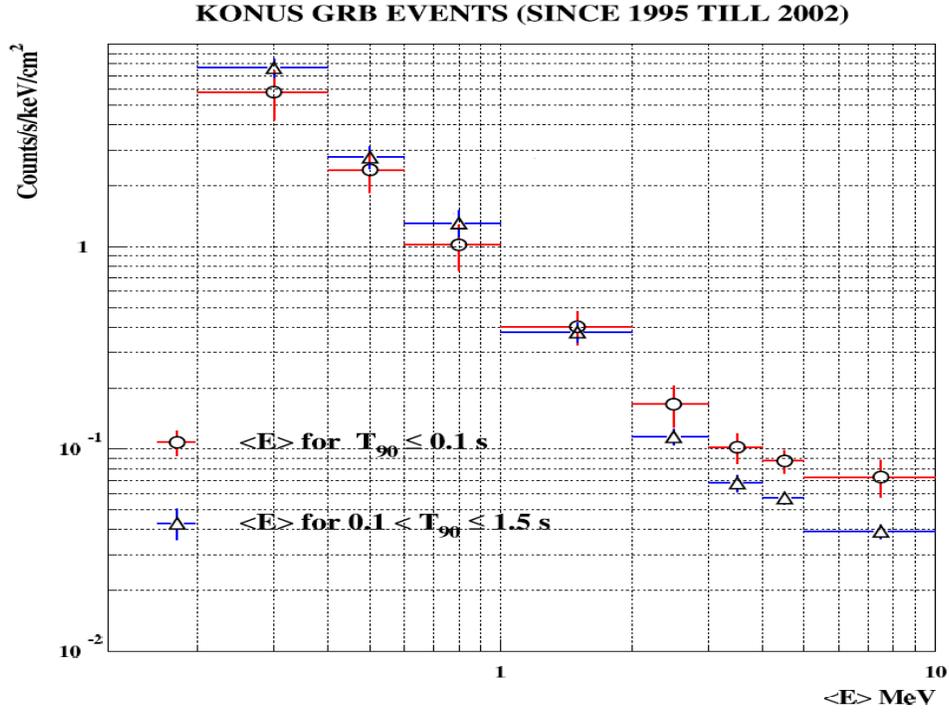

Fig.7. Mean Energy distribution for KONUS Short and Very Short Bursts [5].

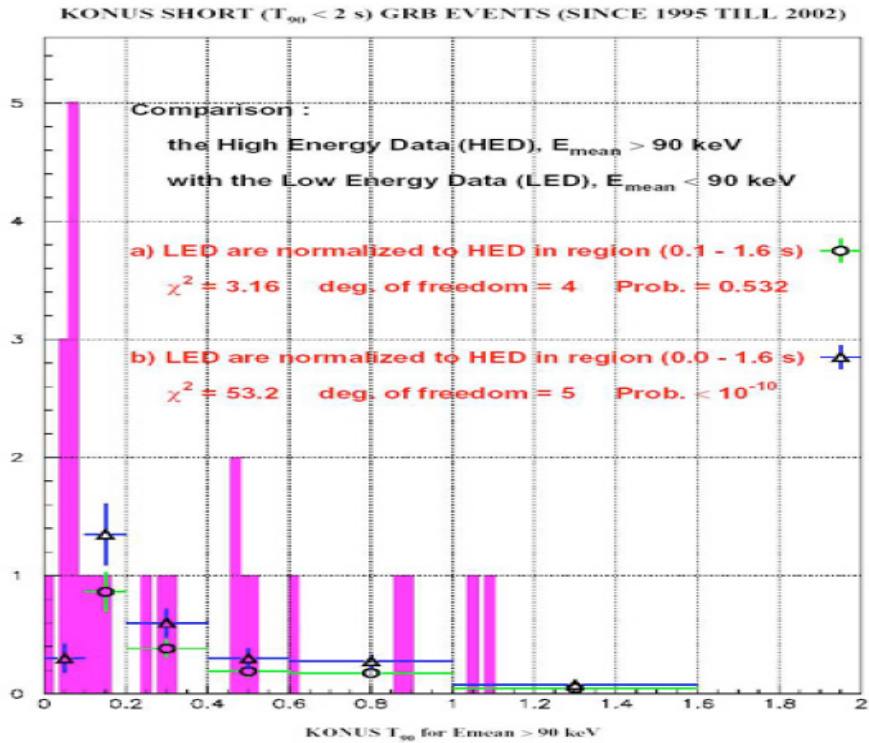

Fig.8. KONUS data with different cuts on the average photon energy ‹$E_\gamma$› [9].

## 6. Afterglows

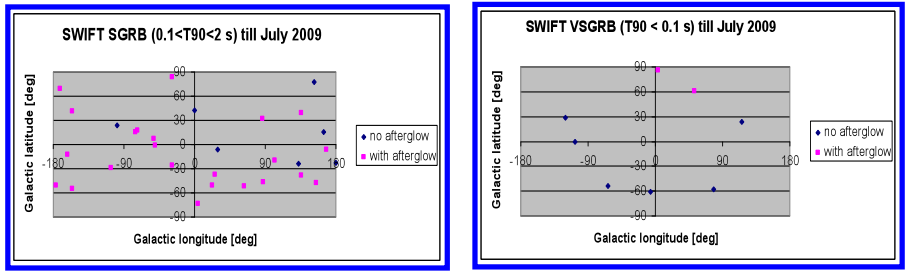

FIG. 9. The SWIFT VSGRB and SGRB angular distribution in Galactic coordinates.

In SWIFT VSGRB sample we observe 25% bursts with afterglows, but in the SGRB sample (0.1 < $T_{90}$ < 2 s) we notice 78% bursts with afterglows. We can interpret it that in VSGRB sample there is about one third of bursts, the tail of the basic SGRB distribution, deriving from the merger of compact objects in binary system (neutron stars or black holes): NS-NS and NS-BH. So two third of VSGRBs are different from SGRBs and can be of new origin, which could be consistent with PBH evaporation.

## 7. Time profile of rising part of BATSE VSGRBs is in agreement with evaporation PBH.

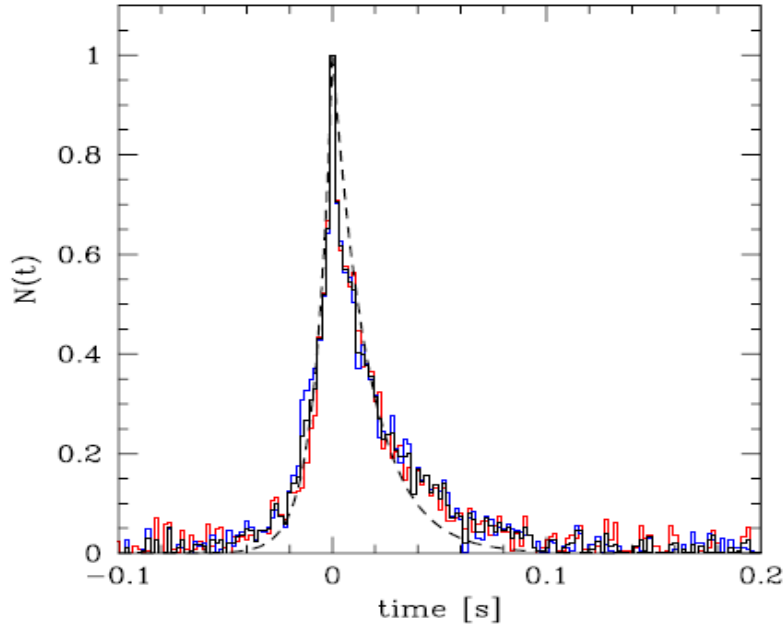

Fig.10. Composite burst profiles for all VSGRB (black line), for bursts from Galactic Anticenter region (red line) and for bursts from outside that region (blue line). The analytical fit (dashed line) is given by Eq.1. Better fit for the decay part is provided by Ryde & Svensson function (Eq. 2) [10].

For all BATSE VSGRBs the background subtracted lightcurve is normalized by its maximum value and shifted, so the maximum is at t = 0. The resulting composite profile is shown in Fig.10.

The profile indicates significant asymmetry with the timescale of $t_{rise}$ = 0.00079 s for exponential rise and of $t_{fall}$ = −0.0171 s for exponential decay time (Eq. 1).

$$f(t) = \exp((t − t_0)/t_{rise}) \text{ for } t < t_0$$
$$= \exp(−(t − t_0)/t_{fall}) \text{ for } t > t_0 \quad (1)$$

If we use evaporation PBH model, under some assumptions [10], the fit to the rising part of the profile is good. The simple evaporation formula does not give the time profile after the peak; it needs some modifications [10]. Better fit to the decay phase, however, is provided by the function

$$f(t) = 1/(1 + t/\tau)^n \quad (2)$$

with index $n$ = 2.62 and the decay timescale $\tau$ = 0.0319 s.
If we supposed the phase transition in final state of evaporation – what means "explosion" with upper limit of particles` and γ` energy, then we should expect modification of the final state of Hawking evaporation process. The energy of the phase transition defines the temperature of the fireball, what denominates total mass evaporated. If energy is lower the mass of fireball is bigger.

## 8. Conclusions

- We observe the time $T_{90}$ distribution of all GRBs detected by 7 experiments and conclude the small surplus of VSGRB ($T_{90}$ < 100 ms) in every experiment, not significant statistically for one detector but meaningful f or all together. BATSE VSGRBs coming from Galaxy Anticenter direction show 4 σ excess for $T_{90}$ < 0.1 s. Also KONUS data with $<E_\gamma>$ > 90 keV gives unexpected excess of VSGRBs in comparison with events with $<E_\gamma>$ < 90 keV. All these signals suggest to treat VSGRB as a new class of events connected with a new origin of the bursts.

- Two independent analyses of BATSE GRB data confirm significant (4 σ) anisotropy in the galactic angular distribution of BATSE VSGRB events. It suggests local origin of these VSGRBs when all the others GRBs have cosmological origin. Also V/Vmax distribution for BATSE VSGRB events indicates its local distance production. This confirms also distinct origin of these events.

- VSGBBs have more energetic γ-ray burst than other GRBs with longer duration (KONUS), see Fig. 7 and Fig. 8. Such effect we would expect for PBH evaporation.

- We observe small number of afterglows in SWIFT VSGRB sample (25%), in contrast with the noticeable afterglow frequency in SGRB sample (78%). We can interpret that in VSGRB sample there is about one third of bursts, (the tail of the basic SGRB distribution, deriving from the merger of compact objects in binary system: NS-NS and NS-BH). So two third of VSGRBs are different from SGRBs and can be of new origin, which could be consistent with PBH evaporation.

- Time profile of rising part BATSE VSGRBs is in agreement with evaporation PBH model. We are not sure the theory of PBH evaporation can be applied [11]. If we supposed the phase transition in final state of evaporation – what means "explosion" with upper limit of particles` and γ` energy, then we should expect modification of the final state of Hawking evaporation process. The energy of the phase transition defines the temperature of the fireball, what denominates total evaporated mass. If energy is lower the mass of fireball is bigger.

All these information suggest that VSGRBs create a new class of GRBs and also suggest that these events can be treated in the majority of cases as a result of PBH evaporation ("explosion").